# The rapid formation of macromolecules in irradiated ice of protoplanetary disk dust traps

**Author list:** Niels F.W. Ligterink[1,2,*], Paola Pinilla[3], Nienke van der Marel[4], Jeroen Terwisscha van Scheltinga[4,5], Alice S. Booth[4,6], Conel M. O'D. Alexander[7], My E.I. Riebe[8]

**Affiliations:**

[1]Physics Institute, Space Research and Planetary Sciences, University of Bern, Sidlerstrasse 5, CH-3012 Bern, Switzerland

ORCID: 0000-0002-8385-9149

[2]Faculty of Aerospace Engineering, Delft University of Technology, Delft, The Netherlands

*email: niels.ligterink@tudelft.nl

[3]Mullard Space Science Laboratory, University College London, Holmbury St Mary, Dorking, Surrey RH5 6NT, UK
ORCID: 0000-0001-8764-1780

[4]Leiden Observatory, Leiden University, P.O. box 9513, 2300 RA Leiden, The Netherlands

ORCID: 0000-0003-2458-9756

[5]Laboratory for Astrophysics, Leiden Observatory, Leiden University, PO Box 9513, 2300 RA Leiden, The Netherlands

ORCID: 0000-0002-3800-9639

[6]Center for Astrophysics, Harvard & Smithsonian, 60 Garden St, Cambridge, MA 02138, USA

ORCID: 0000-0003-2014-2121

[7]Earth and Planets Laboratory, Carnegie Institution for Science, 5241 Broad Branch Road NW, Washington, DC 20015, USA

ORCID: 0000-0002-8558-1427

[8]Institute of Geochemistry and Petrology, Eidgenössische Technische Hochschule Zürich, 8092 Zürich, Switzerland

ORCID: 0000-0002-2098-9587

**Abstract**

Organic macromolecular matter is the dominant carrier of volatile elements such as carbon, nitrogen, and noble gases in chondrites – the rocky building blocks from which Earth formed. How this macromolecular substance formed in space is unclear. We show that its formation could be associated with the presence of dust traps, which are prominent mechanisms for forming planetesimals in planet-forming disks. We demonstrate the existence of heavily irradiated zones in dust traps, where small frozen molecules that coat large quantities of microscopic dust grains could be rapidly converted into macromolecular matter by receiving radiation doses of up to several 10s of eV molecule$^{-1}$ year$^{-1}$. This allows for the transformation of simple molecules into complex macromolecular matter within several decades. Up to roughly 4% of the total disk ice reservoir can be processed this way and subsequently incorporated into the protoplanetary disk midplane where planetesimals form. This finding shows that planetesimal formation and the production of organic macromolecular matter, which provides the essential elemental building blocks for life, might be linked.

**Introduction**

Organic macromolecular matter likely supplied the terrestrial planets with most of their carbon, nitrogen, and noble gases[1]. In chondrites, this material is often called Insoluble Organic Matter (IOM), where it is the dominant carrier of these volatile elements[2]. It has even been suggested that organic macromolecular matter directly contributed to the emergence of life[3]. Similarities in the elemental compositions of chondritic IOM and refractory organic matter in comets, as well as large deuterium enrichments, indicate that there is a genetic relationship between the two materials[4]. A genetic relationship is, perhaps, not so surprising since the formation regions of some chondrites were probably well beyond the orbit of Jupiter[5]. Hence, refractory organic matter was distributed over large radial distances in the proto-Solar Nebula[6]. Detailed characterization of IOM has provided constraints for potential formation mechanisms, but there is still no consensus about whether this material formed in the interstellar medium, the proto-Solar Nebula, or in planetesimals by polymerization of simpler precursors[2]. With a new model, we demonstrate that macromolecular organic matter resembling that of IOM in chondrites[2] and refractory organics in comets[6] could rapidly form in dust traps in the proto-Solar Nebula, via radiation-driven ice chemistry. Large quantities of ice, up to roughly 4% of the total disk ice reservoir in our model, could be processed in this way and later incorporated into planetesimals. Our findings suggest a link between the mechanism that forms planetesimals and the chemical processes that determined their macromolecular and volatile element budgets.

Due to the steady influx of meteorites to Earth, IOM is the best-studied of all extraterrestrial macromolecular organic matter. Its structure is characterized by small (poly)cyclic aromatic units, linked by short, highly branched aliphatic chains, furan/ether bonds, and additional functional groups, such as ketones and carboxyls[7,8]. Deuterium and $^{15}$N enhancements in bulk, and even larger ones in individual grains, require cold (<20 K) and radiation-rich environments where isotopes are readily fractionated, and simple molecules are frozen onto dust grains[9]. Subsequent radiation of frozen organics could enhance deuterium enrichments[10], while heating and aqueous alteration in the parent bodies may have reduced them[11]. IOM contains radicals in its structure[12,13] that may have been generated during irradiation of organic molecules[14]. IOM also contains high concentrations of noble gases which are isotopically fractionated and depleted in light noble gases compared to the Solar composition[15], which is suggestive of a thermal loss mechanism starting at low temperatures (<100 K).

These observations suggest that the formation of IOM and refractory organics in comets involves the irradiation of frozen carbon-bearing molecules. Laboratory studies of particle and UV irradiation of ice films and frozen organic molecules have demonstrated that these molecules can be converted into macromolecular matter[16–19], although mixing ratios in the experiments are not always realistic with respect to the ice found in interstellar environments and protoplanetary disks. Radiation doses of several 100s up to 1000 eV molecules$^{-1}$ are required for the conversion, which is much higher than what molecules in the solid phase typically experience during the star- and planet-formation cycle. UV and cosmic-ray doses received by ice-coated grains in dark clouds and protoplanetary disk midplanes at most reach several 10s of eV molecule$^{-1}$ (~$10^{-7}$ eV molecule$^{-1}$ year$^{-1}$)[20]. Radiation doses received by ice-coated dust grains in the outer layers of a protoplanetary disk, which are directly irradiated by the protostar, are similarly moderate and only sufficient to produce much smaller and typically solvent-soluble organic molecules (SOM, such as amino acids, sugars, polyoxymethylene, and other polymers)[21]. Until now, it was unknown where large quantities of ice and other frozen materials could be heavily irradiated to form organic macromolecular matter during the star- and planet-formation sequence.

Following the first detection of a dust trap in the protoplanetary disk IRS48[22], our understanding of planetesimal formation has been revolutionized[23]. Dust traps are localized pressure bumps in protoplanetary disks where the radial drift of dust is reduced or stopped, and material piles up[24]. Grains in these traps can efficiently grow to form pebbles and pebbles may be further concentrated by the streaming instability to form planetesimals[23,25]. Dust traps are regularly found in protoplanetary disks with different properties (e.g., disk ages from < 1 to 10 Myr) and in the form of rings and crescents in millimeter dust continuum observations with ALMA[23]. They are thought to have played a fundamental role in the formation of the Solar System, as they were likely present in the primordial disk, even beyond the current orbit of Uranus[26]. Based on their distinct isotopic compositions, the different carbonaceous chondrite groups that host IOM have been suggested to have collected in a dust trap that formed in the first 2-4 Myr[27].

Dust traps have been observationally shown to contain large amounts of ice-associated molecules that were likely inherited from the molecular cloud stage[28]. $H_2CO$, $CH_3OH$, and $CH_3OCH_3$ have recently been detected in the gas phase in a region that is co-spatial with dust traps[29–31]. The detection of these molecules can be explained by the vertical transport of ice-coated grains and thermal desorption in the warmer surface layers, whereas in disks without dust traps these molecules remain locked up in ice-coated grains, thus remaining undetectable with ALMA[32]. Hence, vertical transport of ice to irradiated zones of dust traps potentially provides favourable conditions for the formation of organic macromolecules.

In this study, we use the output of a state-of-the-art dust evolution model of a protoplanetary disk with a dust trap[33] and calculate the UV radiation dose rate in ice-coated grains throughout the disk. While we knew of radiation-heavy and ice-rich interstellar environments, prior to our work there was no known environment in a planet-forming disk where high radiation doses and large ice reservoirs (on the percent level) coincided, as most of the ice-coated grains were thought to be well-shielded from radiation. We demonstrate the existence of heavily irradiated regions in dust traps, where large ice reservoirs can rapidly be transformed into organic macromolecular matter such as IOM.

**Results**

The dust and gas distributions predicted by the dust evolution model for a 1 Myr old protoplanetary disk with a dust trap at 45 au are shown in **Fig. 1.A** and **1.B** (details in the **Methods**). The dust-to-gas ratio exceeds unity in the dust trap of this model, which is a sign that planetesimal formation ensues. The modelled dust grain sizes range from $10^{-7}$ to $10^{-1}$ m, but grains smaller than several tens of μm dominate at elevated heights in the dust trap (**Extended Data Figure 1.**). The assumption is made that each dust grain is covered by 100 monolayers (ML, 1 ML = $10^{15}$ molecules $cm^{-2}$) of frozen molecules. Since we determine the average dose rate per molecule, the composition of the ice is arbitrary, but it can be thought to resemble those of interstellar dust grains, that is, $H_2O$ dominated, with fractions of $CH_4$, $CO$, $CO_2$, $CH_3OH$, and $NH_3$. Alternatively, warm grains can be coated in a layer of non-volatile organic molecules after water ice has sublimated[34]. UV photons emitted by the protostar penetrate throughout the disk (**Fig. 1.C**) and the photon flux is calculated as a vertical radiation field, which is attenuated by the gas and dust throughout the disk. The optical depth is determined following $\tau(r, z) = \int_z^\infty \rho(r, z) \kappa dz$ where the column of material at a radius R from the central star above a height Z above the midplane is multiplied by the average opacity $\kappa$ ($cm^2$ $g^{-1}$). The UV flux is calculated with the Beer-Lambert law $F(r, z) = F_0 e^{-\tau(r,z)}$, where F0 is the number of incoming photons. In the fiducial model, we use $F_0 = 1000$ $G_0$, where $G_0$ is a UV flux of $10^8$ photons $cm^{-2}$ $s^{-1}$. Various models of protoplanetary disks demonstrate that this is a flux that is readily achieved in the surface layers of disks, due to the radiation from the central star[35]. The mean opacity is set to $\kappa = 10$ $cm^2$ $g^{-1}$, which is a value commonly used for protoplanetary disks[36].

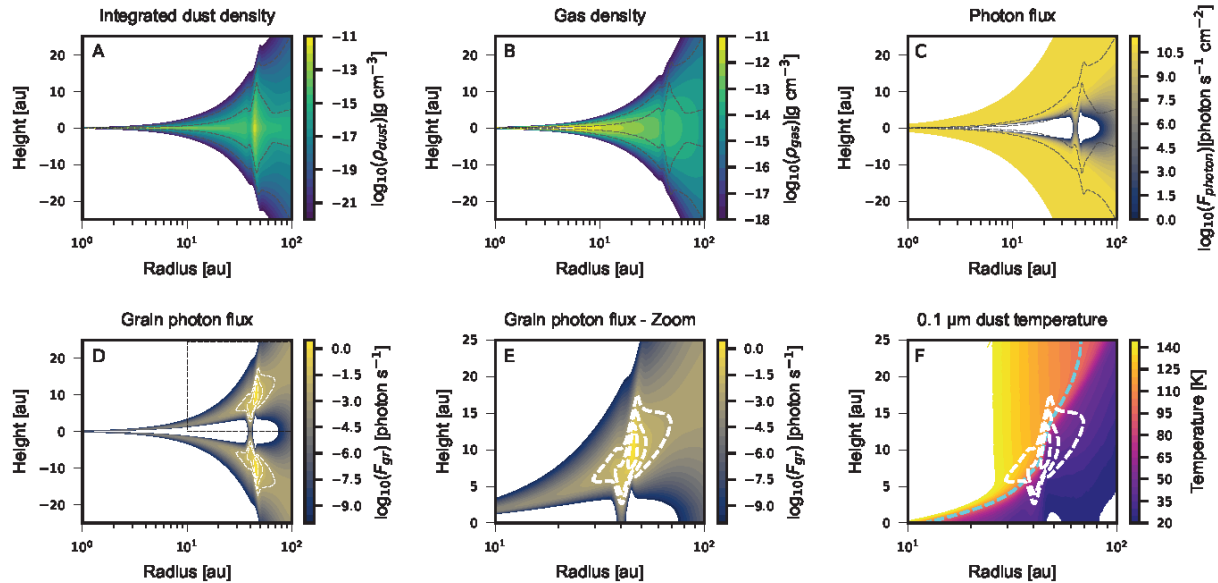

**Fig. 1.** Dust evolution and irradiation model of a protoplanetary disk with a dust trap located at ~45 au. The integrated dust density distribution (**A**), gas density (**B**), photon flux for an opacity of $\kappa = 10$ cm2 g-1 (**C**), grain photon flux (**D,E**), and temperature of the 0.1 μm dust grains (**F**) are shown. The black contours indicate the dust density at $10^{-19}$ (outer) and $10^{-16}$ (inner) g $cm^{-3}$, the hotspot region is depicted by white contours at $10^{-3}$, $10^{-2}$, and $10^{-1}$ times the peak grain photon flux, and the cyan contour indicates the boundary where the grain temperature equals 100 K.

The grain photon flux (product of the photon flux and the grain surface areas) of the fiducial model is shown in Figs. 1.D,E. It reveals that the largest reservoir of ice-coated dust grains receiving the largest photon flux is in the dust trap at Z = ±10 astronomical units (au) and R = 45 au. This is the result of local gas densities being lower near the inner edge of the dust ring or dust trap, while dust densities are large. In comparison, gas densities in the inner disk (taken as r < 10 au in our model) are large (see **Fig. 1.B**) and minor quantities of disk ice receive an appreciable photon flux (see **Fig. 1.D**). In the dust trap at a height of Z = 5 – 20 au, grain temperatures generally exceed ~50 K, while sub-μm grain temperatures surpass even ~100 K in specific regions (see **Fig. 1.F** and **Extended Data Figure 2.**). Therefore, most grains in the dust trap are 'luke-warm' and retain ice films that are dominated by less volatile species (e.g., $H_2O$, $CH_3OH$, PAHs), small amounts of trapped more volatile species, and minor quantities of other organic molecules[34]. The combination of high dust concentrations, moderately elevated temperatures, and radiation makes the dust trap wall a hotspot for radiation-driven ice chemistry, which stimulates the formation of complex molecules[37].

The dust trap hotspot is divided into regions at intervals of $10^{-3}$, $10^{-2}$, and $10^{-1}$ times the peak grain flux ($F_{gr,peak}$). By dividing the incoming photon flux by the amount of ice and assuming an average UV photon energy of 6 eV[38], the average energetic input is determined in eV molecule$^{-1}$ s$^{-1}$ for all molecules in the ice mantle, including $H_2O$. At its peak position, the fiducial model shows that the dose rate is 36 eV molecule$^{-1}$ yr$^{-1}$. A dose of 1000 eV molecule$^{-1}$, which is sufficient to produce macromolecular matter[16–18], is achieved within 30 years. Even in the extended region, where the dose rate decreases to 2 eV molecule$^{-1}$ yr$^{-1}$, this dose is obtained in approximately 500 years. These timescales are well within the expected lifetime of a dust trap[24].

A substantial quantity of ice is contained in the hotspot region, with 2.6%, 2.1%, and 3.5% of the total disk ice reservoir in the three regions, respectively (8% total). Grains are subject to vertical turbulence, which results in the loss of ice-coated grains to the disk atmosphere when grains are stirred upwards. For this model, approximately half of the grains are lost (see **Methods** for details), which means that the other half (4%) of the processed ice reservoir settles to the disk midplane, where it is available as planetesimal-forming material. Because such a large quantity of ice is processed, organic macromolecular matter formation in dust traps can currently best explain the ~6% formation efficiency of IOM and up to 55% for comets as derived from atomic C/Si ratios[2]. The ice-conversion factor is likely higher in a dynamic disk where fresh ice-coated grains are continuously replenished in the dust trap hotspot due to vertical and radial mixing. Conversion efficiencies of pristine ice to organic macromolecular matter have not been determined and therefore the amount of processed ice is an upper limit of produced macromolecular matter. Furthermore, dust trap locations and opacities may differ from disk to disk and will critically influence the dose rates and above-derived conversion fraction. For opacities ranging from 5 – 40 cm$^2$ g$^{-1}$, the average dose rate does not alter significantly for this model (**Fig. 2.A.**), but the amount of processed frozen material increases with decreasing κ, see **Fig. 2.B**. This is the result of photons penetrating deeper into the disk, that is, closer to the midplane, and accessing regions where the concentration of dust grains is larger.

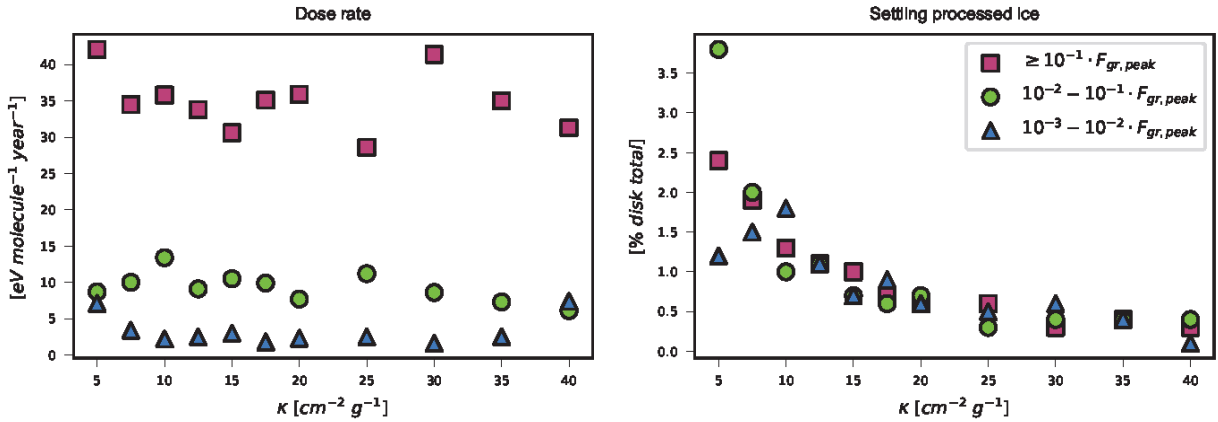

**Fig. 2.** Influence of the opacity on the dose rate and amount of settled processed ice for regions constrained between $10^{-3}$, $10^{-2}$, and $10^{-1}$ the peak grain photon flux. While the dose rate remains constant, the amount of processed ice that can settle back to the midplane increases with decreasing disk opacity.

**Discussion**

The following scenario for the formation of macromolecular matter is proposed (**Fig. 3.**). In dense interstellar clouds, dust grains acquire ice mantles of simple species that are enriched in minor isotopes[39,40]. The ice or dust grain may be enriched in PAHs. Heavy noble gases are efficiently trapped in these ice mantles[41] and moderate grain surface chemistry results in the production of minor quantities of SOM[42]. The protoplanetary disk inherits these ice-coated grains from its host cloud[28]. As grains migrate into the dust trap hotspot, the ice warms up and the UV flux increases, resulting in a rapid conversion of the organic material in the ice mantle into macromolecular matter, by producing and linking complex ice-irradiation products, such as low volatility SOM, and PAH fragments[43–45]. Noble gases in the ice mantles may be trapped in the macromolecular matter during this process[46], and D isotopic signatures could be fractionated during further irradiation due to preferential loss of light isotopes[10,14]. Processed grains are cycled back into the midplane, creating a mixture of pristine ice and macromolecular matter, in line with the fact that chondrites accreted significant amounts of ice[47]. Migration of these mixtures closer to the protostar can result in IOM-rich objects like the asteroid Ryugu[48], whereas staying at these large radii (45 au in this model) preserves it as refractory organics as seen in comets[49]. Since multiple dust traps can occur in a protoplanetary disk, cometary refractory organics and IOM in meteorites and asteroids may be formed at different radii and times. In both cases, the heavy irradiation of grain mantles results in the production of similar macromolecules.

In bulk and at the grain (μm to sub-μm) scale, IOM shows a range of chemical and isotopic characteristics[2,50], which can be explained by this scenario. Disks are turbulent environments[24], creating variations in photon flux and grain surface temperature. Protoplanetary disks can have multiple dust traps at different radii with different shapes and amplitudes, and the flux of radiation from the central star can vary with time[23], which would further enhance the diversity of macromolecular products and their characteristics. For example, deuterium fractionation by irradiation of organic polymers has been demonstrated, where different radiation fluxes will result in different levels of fractionation on top of the inherited signature[10]. The mixing of material that experienced different degrees of alteration of these pathways explains the heterogeneous isotope nature of IOM on (sub)µm scales[50]. Small (sub)µm ice-coated dust grains are preferentially stirred up into the hotspot region (**Extended Data Figure 1.**), but trajectories and residence times may vary from a few to several dozen $\Omega_k^{-1}$ (in this model, $\Omega_k^{-1}$ is ~30-50 years at the hotspot radius)[51]. However, because small grains are continuously re-created in the trap due to the fragmentation of the large particles, there is a continuous population of small grains present for million-year timescales that are exposed to the irradiation needed for IOM formation. This affects the transformation of pristine material into organic macromolecular matter and, in turn, its chemical composition and the texture of the material. For example, nanoglobules, small spherical and often hollow macromolecular grains[52,53], have been suggested to have formed by UV irradiation of ice mantles of (sub)micrometer-sized grains[53], although alternative formation pathways by aqueous alteration have been suggested as well[8].

A prominent question that remains, is whether organic macromolecular matter is formed in a water-rich or -poor environment. Ice mixtures in irradiation experiments that produce IOM analogues contain minor or no water[17,18] and do not match with the known ice compositions of interstellar ice-coated dust grains or comets[34]. Water-rich ice is expected to lead to enhanced $CO_2$ formation and oxygen-rich macromolecular material, inconsistent with the carbon-rich elemental composition of organic macromolecular matter observed in meteorites and comets. From the experimental work conducted thus far, it is unknown whether macromolecular matter can form in water-dominated ice. Instead, its formation may rely on a preceding step where low volatility SOM is formed, water-ice is removed (e.g., by thermal desorption)[42], and followed by heavy irradiation to form complex refractory macromolecules[16,19]. In our model, some of the smallest dust grains, which dominate the dust composition in the hotspot, reach temperatures at which water-ice readily sublimates. If a dust trap is located closer to the protostar, grains of larger sizes will be heated up to higher temperatures and the loss of water-ice becomes more prominent, while organics of low volatility remain. A thermal heating cycle may also have further depleted Ar and Kr relative to Xe, resulting in the low Ar/Xe and Kr/Xe ratios in natural IOM.

To determine the required dose to form organic macromolecular matter, we cite particle irradiation studies[16–18], whereas our model relies on UV radiation. Photons are affected by the optical properties of the ice, whereas penetration depths of particles of keV and larger energies are usually larger than UV photons. These differences may result in different chemical outcomes of ice irradiation. However, experimental products are the same for ice films irradiated with UV photon or ions at similar dose[54]. This makes it plausible that both radiation types will result in a similar type of organic macromolecular matter, as indirectly indicated by the UV irradiation experiments[19].

In the past few years, high-resolution observations of protoplanetary disks have demonstrated that disks have dust traps irrespective of their properties or those of their central star. We demonstrate that hotspots, that is, zones of heavy irradiation and increased dust temperatures, exist in these dust traps, where the icy mantle material on dust grains can rapidly be converted to organic macromolecular matter. Since dust traps are also likely places for efficient planetesimal formation, this suggests that the mechanism by which planetesimals form also produced the macromolecular matter from which the terrestrial planets derive their elemental carbon and nitrogen, which in the case of the Earth contributed to the emergence of life.

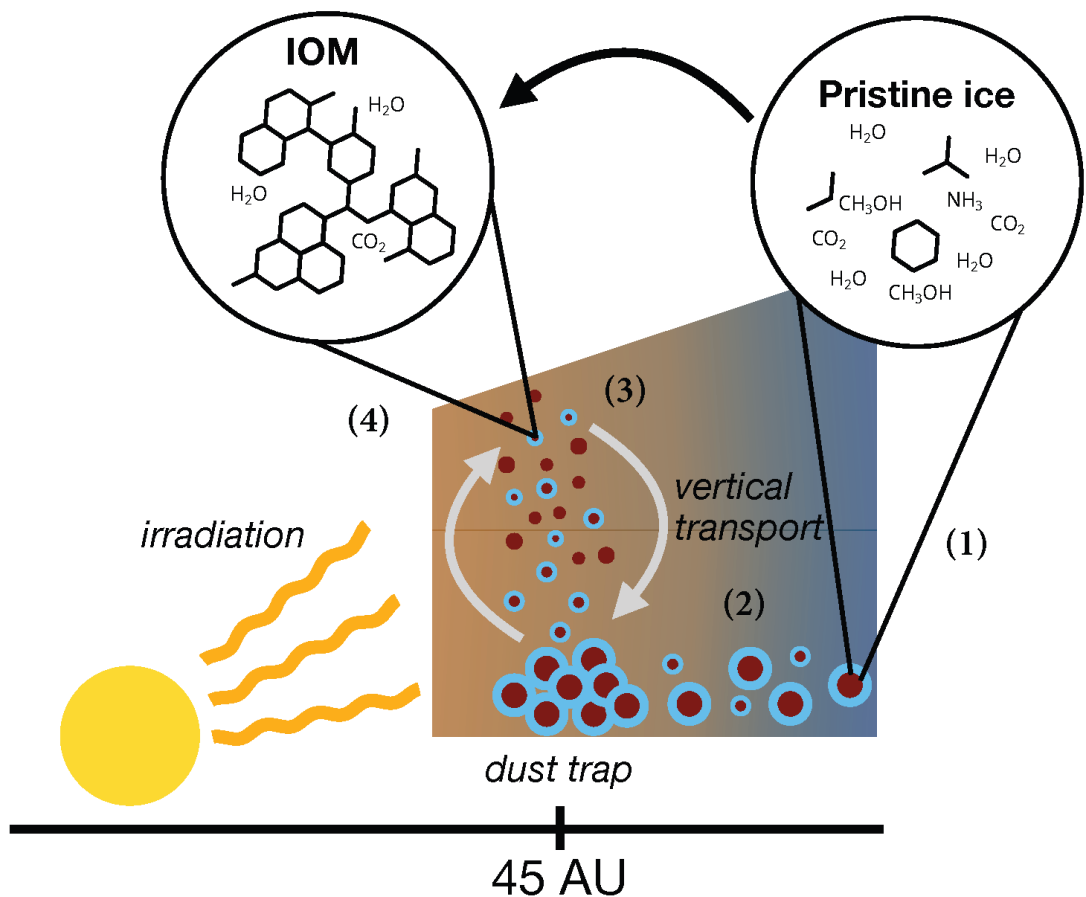

**Fig. 3.** Schematic depiction of the IOM formation scenario. Grains with pristine, simple ice components (1) radially drift into the protoplanetary disk (2) and migrate to the dust trap hotspot through vertical mixing/transport (3), where heavy irradiation of the luke-warm ice results in the formation of organic macromolecular matter (4).

**Methods**

The dust evolution models are performed using the Dustpy code[55]. The disk is assumed to be around a Solar-mass star, with a gas surface density that assumes a critical radius at 80 AU and an initial disk mass of 5% of the Sun[24]. The model assumes a fragmentation velocity of $v_f = 10$ m s$^{-1}$, a gas viscous evolution parameter, radial diffusion, turbulent mixing, and vertical settling/stirring all set to $10^{-4}$. All grains are initially small between 0.1-1 microns in size, with a power law distribution as $n(a) \propto a^{-3.5}$. The models include the dust growth and dynamics of particles. The model output gives the density distribution of the dust grains ($\rho_{dust}$, **Fig. 1.A**) and of the gas ($\rho_{gas}$, **Fig. 1.B**) at 1 Myr of evolution. The data is portrayed in a radial grid from 1 to 300 au (only 1 to 100 au shown in **Fig. 1**) with 300 logarithmically spaced cells. A logarithmically spaced grain size distribution between the minimum grain size of 0.1 micron to 0.1 m in 127 steps is used. A gap centred at 40 AU is assumed in the disk[33], which yields a pressure bump at around 45 AU. Smaller and lighter dust grains are more easily stirred up in the disk and therefore dominate the particle sizes at greater disk heights, whereas larger particles concentrate around the disk midplane (**Extended Data Figure 1.**)

From the dust density distribution, the grain surface area is calculated by assuming that all grains are spherical and have a density of $\rho_{grain}$ = 1.65 g cm$^{-3}$. This allows us to calculate the particle mass for each grain size and convert the dust density (g cm$^{-3}$) into a particle density (n cm$^{-3}$). Next, the particle density is multiplied by the grain area to find the total grain area per volume, which gives:

$$A_{grain} = \frac{(3 \cdot \rho_{dust})}{(\rho_{grain} \cdot r_{grain})} \qquad (1)$$

where $r_{grain}$ is the grain size (that is, the radius). Using the grain area, the available amount of ice is calculated by assuming that each grain is covered with 100 monolayers (1 ML = $10^{15}$ molecules cm$^{-2}$) of ice and by multiplying this value with the available grain surface area.

The thermal structure of the disk is modelled with RADMC-3D[56], assuming a vertical grid of 180 cells over a semicircle following Z = R · cos(θ), where Z is the disk height and R the radius, and following the same procedure as[33]. The grain sizes that dominate the dust trap hotspot have sizes of 0.1 to several tens of μm and generally have temperatures greater than 50 K. However, only sub-μm grains are heated above 100 K, and only in specific regions of the hotspot (**Extended Data Figure 2.**).

The number of photons throughout the disk is calculated with the Beer-Lambert law (see main text). The impinging photon flux is fixed to $F_0$ = 1000 $G_0$ ($G_0$ = $10^8$ photons cm$^{-2}$ s$^{-1}$), but we note the (grain) photon flux throughout the disk scales linearly with the value chosen for $F_0$. Therefore, the dose rate can be scaled in the same way. The number of photons absorbed by the mantle depends on its thickness and the UV photon absorption cross-section. Assuming the mantle consists entirely of water ice, then a film of 100 ML does not fully absorb the received flux[57]. The simplified division of the photon energy input by the column density to yield eV molecule$^{-1}$ s$^{-1}$ is therefore a rough assumption. However, since the photons that penetrate the ice mantle still hit the underlying grain, we assume that the energy is contributed to the overall system and therefore the simplified division holds.

Gravitational attraction causes grains to settle in the disk midplane, while turbulent stirring can move particles towards or away from the midplane. The velocities of these processes can be calculated[58] and in turn, be used to assess how the dust is vertically distributed.

The equation to determine the velocity with which dust settles to midplane is given as:

$$v_{sett} = z\, \Omega_k\, ST, \qquad (2)$$

where z is the vertical height in meter and $\Omega_k$ is the Keplerian frequency:

$$\Omega_k = \sqrt{\frac{G\, M_\odot}{R^3}}, \qquad (3)$$

with G the gravitational constant, $M_\odot$ the Solar mass in kg, and R the radius in m, and $ST$ the Stokes parameter, calculated following:

$$ST = \frac{r_{grain}\, \rho_{grain}}{\Sigma_g} \frac{\pi}{2} \qquad (4)$$

where $\Sigma_g$ the gas surface density.

To calculate the vertical velocity, we assume a balance between settling and vertical turbulence and that the later velocity is as the relative velocities due to turbulence of particles of similar size, and use the following equation:

$$v_{stir} = \sqrt{\frac{3\,\alpha}{ST + ST^{-1}}}\, c_S \qquad (5)$$

is used, where α is the gas viscous evolution parameter (α = $10^{-4}$) and $c_S$ is the isothermal sound velocity:

$$c_S = \sqrt{\frac{k_B T_{gas}}{\mu_{gas} m_{proton}}}, \qquad (6)$$

with $T_{gas}$ the gas temperature ($T_{gas}$ = 50 K, the average dust temperature at the trap location and assuming $T_{gas} = T_{dust}$), $\mu_{gas}$ the mean molecular mass of the gas ($\mu_{gas}$ = 2), and $m_{proton}$ the proton mass in kg. Note that in the disk surface, where our calculations are relevant, the dust densities are dominated by the small grains, so the dust particles have similar Stokes numbers, which endorse the use of **Eq (5)** for the stirring velocities.

For the model results used in this work, $v_{stir} \gg v_{sett}$ for the µm-sized particles that reside at **a** greater height in the dust trap. Therefore, stirring determines which direction the material migrates. Since the vertical motion of particles can point in two directions, namely away from the midplane (up) or towards the midplane (down), we assume that the likelihood of up- or downward motion is equal. This means that 50% of the material in the dust trap hotspot goes to the disk atmosphere and 50% moves towards the midplane. We note that this situation holds for a static snapshot of the disk model, but in a dynamic and evolving environment, the loss and settle fractions may be different.

**Data availability:** The datasets used and generated during this study are available in the Zenodo repository: https://zenodo.org/records/11953364

**Code availability:** The code used for this study is available in the Zenodo repository: https://zenodo.org/records/11953364

**Acknowledgements:** The authors thank the three reviewers for their valuable input and insights, which helped strengthen the paper. The authors thank Eva G. Bøgelund and Maria N. Drozdovskaya for insightful discussions. N.F.W.L. is supported by the Swiss National Science Foundation (SNSF) Ambizione Grant 193453. P.P. acknowledges the support from the UK Research and Innovation (UKRI) under the UK government's Horizon Europe funding guarantee from ERC (under grant agreement No 101076489). A.S.B. is supported by a Clay Postdoctoral Fellowship from the Smithsonian Astrophysical Observatory. M.E.I.R. is supported by the SNSF Ambizione grant 193331.

**Author contributions:** N.F.W.L. developed the model with support from P.P. Analysis and interpretation of the model results were performed by N.F.W.L., P.P., N.v.d.M., J.T.v.S., A.B., C.M.O'D.A., and M.E.I.R. All authors contributed to writing and review of the manuscript.

**Competing interests:** The authors declare no competing interests.

**Tables:** Not applicable.

**Figure Legends/Captions (for main text figures)**

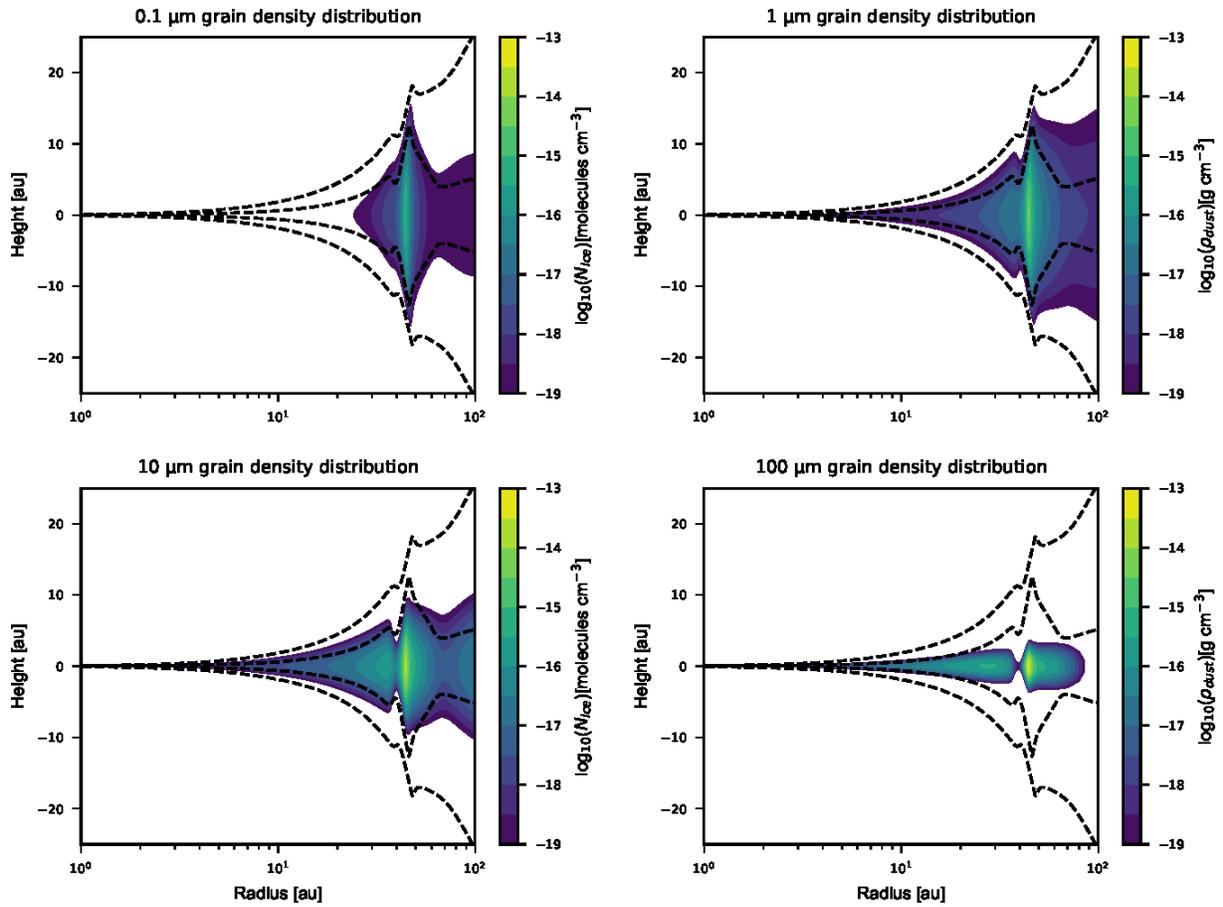

**Extended Data Figure 1.** Dust density distribution for grains of 0.1, 1, 10, and 100 μm in size. The black dashed lines indicate the total dust density distribution contours at $10^{-19}$ and $10^{-16}$ g cm$^{-3}$.

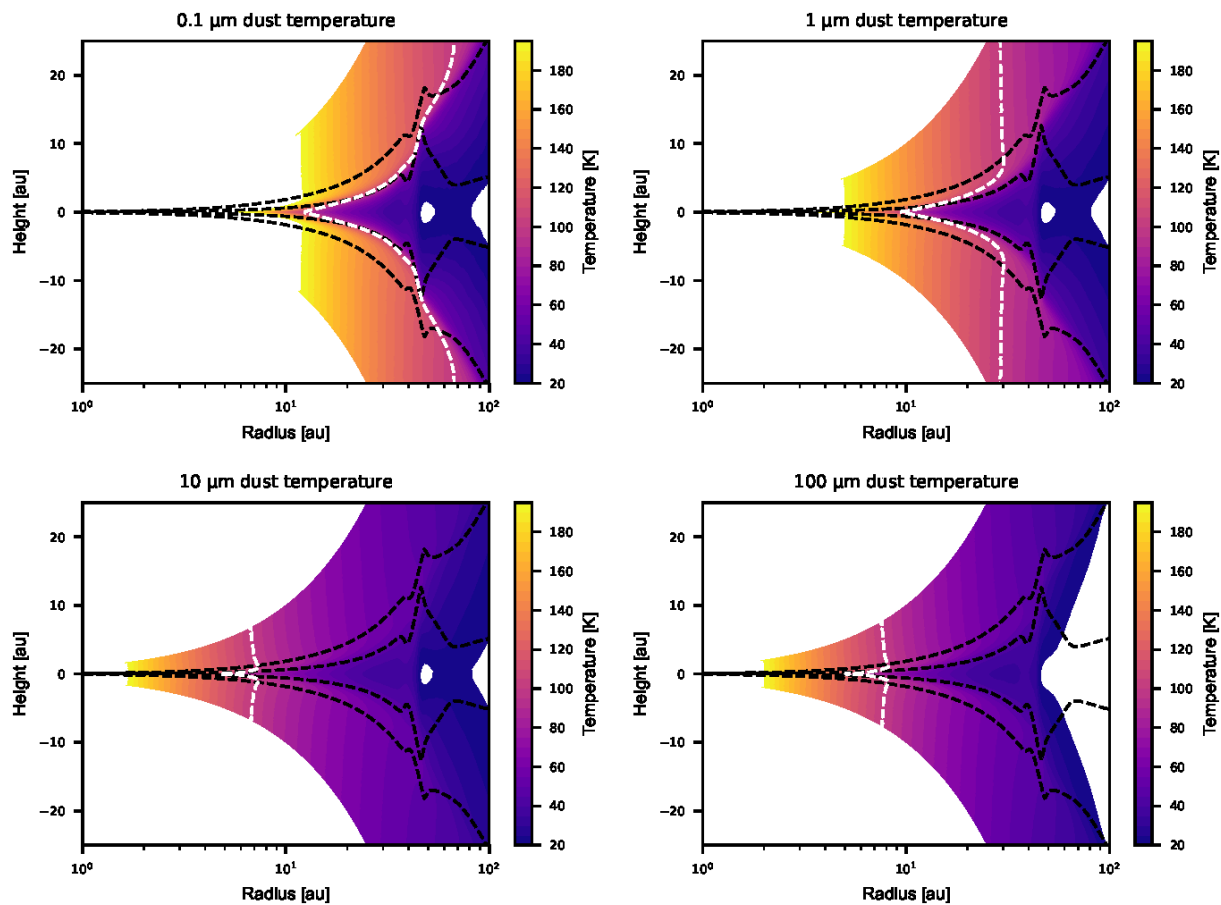

**Extended Data Figure 2.** Dust temperatures for grains of 0.1, 1, 10, and 100 μm in size. The white dashed lines indicate the 100 K temperature contour, while the black dashed lines indicate the total dust density distribution contours at $10^{-19}$ and $10^{-16}$ g cm$^{-3}$.